\documentclass[journal]{IEEEtran}
\usepackage{amsmath}
\usepackage{mathrsfs}
\usepackage{amsfonts}
\usepackage{textcomp}
\usepackage{diagbox}
\usepackage{multirow}
\usepackage{float}
\usepackage{colortbl}
\usepackage{diagbox}
\usepackage{hhline}
\usepackage{comment}
\usepackage{subfig}
\usepackage[table]{xcolor}
\usepackage{tabularx,booktabs}
\usepackage{siunitx}
\usepackage{collcell}
\usepackage{hhline}
\usepackage{pgf}
\usepackage[noadjust]{cite}
\usepackage[labelsep=period]{caption}
\usepackage[font=footnotesize]{caption}

\ifCLASSINFOpdf
\else
\fi
\begin{document}
\title{Real-Time Radar-Based Gesture Detection and Recognition Built in an Edge-Computing Platform}

\author{Yuliang~Sun,~\IEEEmembership{Student Member,~IEEE,}
        Tai~Fei,~\IEEEmembership{Member,~IEEE,}
        Xibo~Li,
        Alexander~Warnecke,
        Ernst~Warsitz,
        and~Nils~Pohl,~\IEEEmembership{Senior~Member,~IEEE}% <-this % stops a space

        \thanks{A video is available on https://youtu.be/IR5NnZvZBLk}
		\thanks{This article will be published in a future issue of IEEE Sensors Journal.}
        \thanks{DoI: 10.1109/JSEN.2020.2994292}

%		\thanks{1558-1748~\copyright2020 IEEE. Personal use of this material is permitted.  Permission from IEEE must be obtained for all other uses, in any current or future media, including reprinting/republishing this material for advertising or promotional purposes, creating new collective works, for resale or redistribution to servers or lists, or reuse of any copyrighted component of this work in other works.}
}

% <-this % stops a space


\markboth{Accepted by IEEE Sensors Journal}%
{Shell \MakeLowercase{\textit{et al.}}: Bare Demo of IEEEtran.cls for IEEE Journals}



\IEEEpubid{\begin{minipage}{\textwidth}\ \\[12pt] \centering
		1558-1748~\copyright2020 IEEE. Personal use of this material is permitted.  Permission from IEEE must be obtained for all other uses, in any current or future media, including reprinting/republishing this material for advertising or promotional purposes, creating new collective works, for resale or redistribution to servers or lists, or reuse of any copyrighted component of this work in other works.
\end{minipage}}

\maketitle

\begin{abstract}
In this paper, a real-time signal processing framework based on a 60 GHz frequency-modulated continuous wave (FMCW) radar system to recognize gestures is proposed. In order to improve the robustness of the radar-based gesture recognition system, the proposed framework extracts a comprehensive hand profile, including range, Doppler, azimuth and elevation, over multiple measurement-cycles and encodes them into a feature cube. Rather than feeding the range-Doppler spectrum sequence into a deep convolutional neural network (CNN) connected with recurrent neural networks, the proposed framework takes the aforementioned feature cube as input of a shallow CNN for gesture recognition to reduce the computational complexity. In addition, we develop a hand activity detection (HAD) algorithm to automatize the detection of gestures in real-time case. The proposed HAD can capture the time-stamp at which a gesture finishes and feeds the hand profile of all the relevant measurement-cycles before this time-stamp into the CNN with low latency. Since the proposed framework is able to detect and classify gestures at limited computational cost, it could be deployed in an edge-computing platform for real-time applications, whose performance is notedly inferior to a state-of-the-art personal computer. The experimental results show that the proposed framework has the capability of classifying 12 gestures in real-time with a high $F_1$-score.
\end{abstract}


% \fi
\IEEEpeerreviewmaketitle

\section{Introduction}
\label{introduction}
\IEEEPARstart{R}{ADAR} sensors are being widely used in many long-range applications for the purpose of target surveillance, such as in aircrafts, ships and vehicles \cite{chen2006micro, hasch2012millimeter}. Thanks to the continuous development of silicon techniques, various electric components can be integrated in a compact form at a low price \cite{hasch2012millimeter,pohl2012ultra}. Since radar sensors become more and more affordable to the general public, numerous emerging short-range radar applications, e.g., non-contact hand gesture recognition, are gaining tremendous importance in efforts to improve the quality of human life \cite{gurbuz2019radar, le2019radar}. Hand gesture recognition enables users to interact with machines in a more natural and intuitive manner than conventional touchscreen-based and button-based human-machine-interfaces \cite{gu2019motion}. For example, Google has integrated a 60 GHz radar into the smartphone Pixel 4, which allows users to change songs without touching the screen \cite{pixel}. What's more, virus and bacteria surviving on surfaces for a long time could contaminate the interface and cause people's health problems. For instance, in 2020, tens of thousands of people have been infected with COVID-19 by contacting such contaminate surfaces \cite{kampf2020persistence}. Radar-based hand gesture recognition allows people to interact with the machine in a touch-less way, which may reduce the risk of being infected with virus in a public environment. Unlike optical gesture recognition techniques, radar sensors are insensitive to the ambient light conditions; the electromagnetic waves can penetrate dielectric materials, which makes it possible to embed them inside devices. In addition, because of privacy-preserving reasons, radar sensors are preferable to cameras in many circumstances \cite{sun2018gesture}. Furthermore, computer vision techniques applied to extract hand motion information in every frame are usually not power efficient, which is therefore not suitable for wearable and mobile devices \cite{lien2016soli}.

\IEEEpubidadjcol
Motivated by the benefits of radar-based touch-less hand gesture recognition, numerous approaches were developed in recent years. The authors in \cite{sun2018gesture,amin2018hand,kim2016hand} extracted physical features from micro-Doppler signature \cite{chen2006micro} in the time-Doppler-frequency (TDF) domain to classify different gestures. Li \textit{et al.} \cite{li2017sparsity}
extracted sparsity-based features from TDF spectrums for gesture recognition using a Doppler radar. In addition to Doppler information of hand gestures, the Google Soli project  \cite{lien2016soli,wang2016interacting} utilized the range-Doppler (RD) spectrums for gesture recognition via a 60 GHz frequency-modulated continuous wave (FMCW) radar sensor. Thanks to the wide available bandwidth (7 GHz), their systems could recognize fine hand motions. Similarly, the authors in \cite{wang2019ts,choi2019short,hazra2019short} also extracted hand motions based on RD spectrums via an FMCW radar. In \cite{skaria2019hand,sun2019automatic}, apart from the range and Doppler information of hand gestures, they also considered the incident angle information by using multiple receive antennas to enhance the classification accuracy of their gesture recognition system. However, none of the aforementioned techniques exploited all the characteristics of a gesture simultaneously, i.e., range, Doppler, azimuth, elevation and temporal information. For example, in \cite{sun2018gesture,lien2016soli,amin2018hand,kim2016hand,li2017sparsity,wang2016interacting,wang2019ts,choi2019short}, they could not differentiate
two gestures, which share similar range and Doppler information. This restricts the design of gestures to be recognized.

In order to classify different hand gestures, many research works employed artificial neural networks for this multi-class classification task. For example, the authors in \cite{kim2016hand,skaria2019hand,sun2019automatic,zhang2019u} considered the TDF spectrums or range profiles as images and directly fed them into a deep convolutional neural network (CNN). Whereas, other research works \cite{zhang2018latern,wang2019ts,wang2016interacting} considered the radar data over multiple measurement-cycles as a time-sequential signal, and utilized both the CNNs and recurrent neural networks (RNNs) for gesture classification. The Soli project \cite{wang2016interacting} employed a 2-dimensional (2-D) CNN with a long short-term memory (LSTM) to extract both the spatial and temporal features, while the Latern \cite{zhang2018latern,zhang2018riddle} replaced the 2-D CNN with 3-D CNN \cite{ji20123d} followed by several LSTM layers. Because the 3-D CNN could extract not only the spatial but also the short-term temporal information from the RD spectrum sequence, it results in a better classification accuracy than the 2-D CNN \cite{zhu2017multimodal}. However, the proposed 2-D CNN, 3-D CNN and LSTM for gesture classification require huge amounts of memory in the system, and are computationally inefficient. Although Choi \textit{et al.} \cite{choi2019short} projected the range-Doppler-measurement-cycles into range-time and Doppler-time to reduce the input dimension of the LSTM layer and achieved a good classification accuracy in real-time, the proposed algorithms were implemented on a personal computer with powerful computational capability. As a result, the aforementioned radar-based gesture recognition system in \cite{kim2016hand,skaria2019hand,sun2019automatic,zhang2018latern,wang2019ts,wang2016interacting,zhang2019u,choi2019short} are not applicable for most commercial embedded systems such as wearable devices, smartphones, in which both memory and computational power are limited.
\begin{figure*}[h]
	\centering
	\includegraphics[width=18cm]{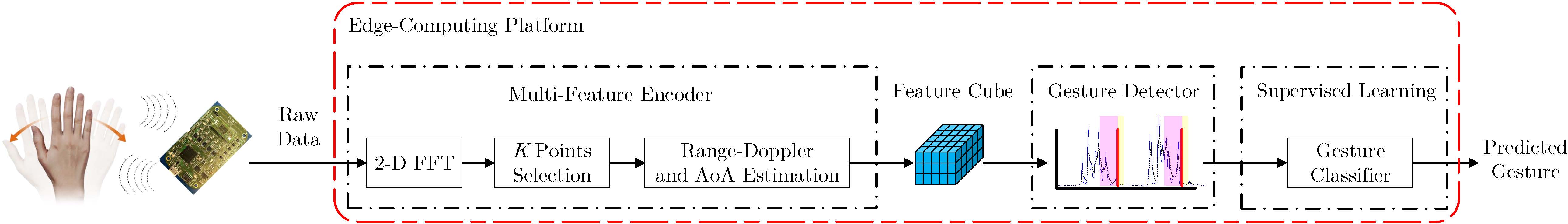}
	\caption{Proposed real-time radar-based gesture recognition framework built in an edge-computing platform.}
	\label{framework}
\end{figure*}

In this paper, we present a real-time gesture recognition system using a 60 GHz FMCW radar in an edge-computing platform. The proposed system is expected to be applied in short-range applications (e.g., tablet, display, and smartphone) where the radar is assumed to be stationary to the user. The entire signal processing framework is depicted in Fig.~\ref{framework}. After applying the 2-dimensional finite Fourier transform (2-D FFT) to the raw data, we select a certain number of points from the resulting RD spectrum as an intermediate step rather than directly putting the entire spectrum into deep neural networks. Additionally, thanks to the L-shaped receive antenna array, the angle of arrival (AoA) information of the hand, i.e., azimuth and elevation, can be calculated. For every measurement-cycle, we store this information in a feature matrix with reduced dimensions. By selecting a few points from the RD spectrum, we reduce the input dimension of the classifier and limit the computational cost. Further, we present a hand activity detection (HAD) algorithm called the short-term average/long-term average (STA/LTA)-based gesture detector. It employs the concept of STA/LTA \cite{vaezi2015comparison} to detect when a gesture comes to an end, i.e., the tail of a gesture. After detecting the tail of a gesture, we arrange the feature matrices belonging to the measurement-cycles, which are previous to this tail, into a feature cube. This feature cube constructs a compact and comprehensive gesture profile which includes the features of all the dominant point scatters of the hand. It is subsequently fed into a shallow CNN for classification. The main contributions are summarized as follows: 
\begin{itemize}
	\item The proposed signal processing framework is able to recognize more gestures (12 gestures) than those reported in other works in the literature. The framework can run in real-time built in an edge-computing platform with limited memory and computational capability.
	\item We develop a multi-feature encoder to construct the gesture profile, including range, Doppler, azimuth, elevation and temporal information into a feature cube with reduced dimensions for the sake of data processing efficiency. 
	\item We develop an HAD algorithm based on the concept of STA/LTA to reliably detect the tail of a gesture.
	\item Since the proposed multi-feature encoder has encoded all necessary information in a compact manner, it is possible to deploy a shallow CNN with a feature cube as its input to achieve a promising classification performance.	  
	\item The proposed framework is evaluated twofold: its performance is compared with the benchmark in off-line scenario, and its recognition ability in real-time case is assessed as well. 	
	\end{itemize}
The remainder of this paper is organized as follows. Section \ref{radar} introduces the FMCW radar system. Section \ref{preprocessing} describes the multi-feature encoder including the extraction of range, Doppler and AoA information. In Section \ref{detection}, we introduce the HAD algorithm based on the concept of the STA/LTA. In Section \ref{supervised}, we present the structure of the applied shallow CNN for gesture classification. In Section \ref{dataset}, we describe the experimental scenario and the collected gesture dataset. In Section \ref{experiment}, the performance is evaluated in both off-line and real-time cases. Finally, conclusions are given in Section \ref{conclusion}.

\section{Short-Range FMCW Radar System}
\label{radar}
Our 60 GHz radar system adopts the linear chirp sequence frequency modulation \cite{kronauge2014new} to design the waveform. After mixing, filtering and sampling, the discrete beat signal consisting of $I_T$ point scatters of the hand in a single measurement-cycle from the $z$-th receive antenna can be approximated as \cite{sun2019high}: 
\begin{equation}
\begin{split}
b^{(z)}(u, v) & \approx \sum_{i=1}^{I_T} a^{(z)}_i \exp \left\{j 2\pi \left(f_{\mathrm{r}i}uT_\mathrm{s} - f_{\mathrm{D}i}vT_\mathrm{c}  \right) \right\},\\ 
u &= 0, \cdots, I_\mathrm{s}-1, \quad v = 0, \cdots, I_\mathrm{c}-1,\\  
\end{split}
\label{fB_multi}
\end{equation}
where the range and Doppler frequencies $f_{\mathrm{r}i}$ and $f_{\mathrm{D}i}$ are given as:
\begin{equation}
f_{\mathrm{r}i} = 2\frac{f_{\mathrm{B}}}{T_\mathrm{c} } \frac{r_i}{c}, \quad f_{\mathrm{D}i} = 2 \frac{v_{\mathrm{r}i}}{\lambda},
\label{frd}
\end{equation}
respectively, $r_i$ and $v_{\mathrm{r}i}$ are the range and relative velocity of the $i$-th point scatter of the hand, $f_{\mathrm{B}}$ is the available bandwidth, $T_\mathrm{c}$ is the chirp duration, $\lambda$ is the wavelength at 60 GHz, $c$ is the speed of light, the complex amplitude $a^{(z)}_i$ contains the phase information, $I_\mathrm{s}$ is the number of sampling points in each chirp, $I_\mathrm{c}$ is the number of chirps in every measurement-cycle, and the sampling period $T_\mathrm{s} = T_\mathrm{c}/I_\mathrm{s}$. 
\begin{figure}[htbp]
	\centering
	\includegraphics[width=7cm]{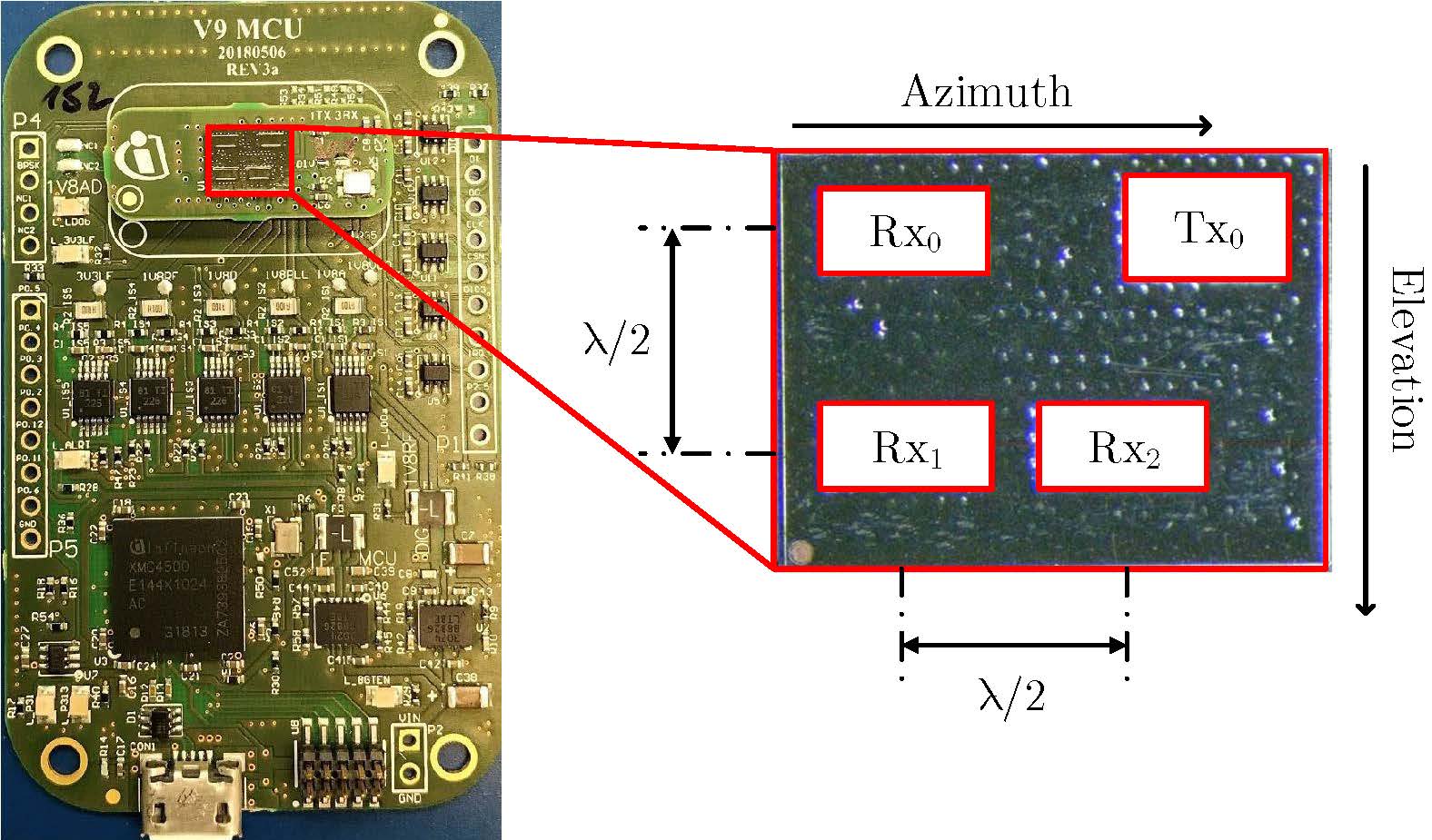}
	\caption{Infineon's BGT60TR13C 60 GHz radar system for gesture recognition. The $\left\lbrace \mathrm{Rx}_z: z = 0,1,2\right\rbrace $ denotes the $z$-th receive antenna. The pair consisting of $\mathrm{Rx}_1$ and $\mathrm{Rx}_0$ is responsible for elevation angle, and the pair consisting of $\mathrm{Rx}_1$ and $\mathrm{Rx}_2$ is used for azimuth angle calculation.}
	\label{evalkit}
\end{figure}
The 60 GHz radar system applied for gesture recognition can be seen in Fig.~\ref{evalkit}. It can also be seen that, the radar system has an L-shaped receive antenna array. To calculate the AoA in azimuth and elevation directions, the spatial distance between two receive antennas in both directions is $d$, where  $d = \lambda/2$.
\section{Multi-Feature Encoder}
\label{preprocessing}
\subsection{2-D Finite Fourier Transform}
A 2-D FFT is applied to the discrete beat signal in \eqref{fB_multi} to extract the range and Doppler information in every measurement-cycle \cite{sun2018two}. The resulting complex-valued RD spectrum for the $z$-th receive antenna can be calculated as:
\begin{equation}
\begin{split}
B^{(z)}(p,q) &= \frac{1}{I_\mathrm{c} I_\mathrm{s}} \sum_{v=0}^{I_\mathrm{c}-1}\sum_{u=0}^{I_\mathrm{s}-1}  \left\lbrace b^{(z)}(u,v) w(u,v)\right\rbrace \\
& \cdot  \exp\left( -j2\pi \frac{p u}{I_\mathrm{s}} \right)\cdot \exp \left( -j2 \pi \frac{qv}{I_\mathrm{c}}\right),\\
p&= 0 , \cdots, I_\mathrm{s}-1, \quad q = 0 , \cdots, I_\mathrm{c}-1,
\end{split}
\label{B1t}
\end{equation}
where $w(u,v)$ is a 2-D window function, $p$ and $q$ are the range and Doppler frequency indexes. The range and relative velocity resolution can be deduced as:
\begin{equation}
\Delta r= c\frac{T_\mathrm{c}}{2 f_{\mathrm{B}}}\cdot \Delta f_\mathrm{r}=\frac{c}{2f_{\mathrm{B}}}, \quad \Delta v_\mathrm{r}= \frac{\lambda}{2} \cdot \Delta f_\mathrm{D}, 
\label{drdv}
\end{equation}
where the range and Doppler frequency resolution $\Delta f_\mathrm{r}$ and $\Delta f_\mathrm{D}$ are $1/T_\mathrm{c}$ and $1/(I_\mathrm{c}T_\mathrm{c})$, respectively. To improve the signal-to-noise ratio (SNR), we sum the RD spectrums of the three receive antennas incoherently, i.e.,
\begin{equation}
RD(p, q) = \sum_{z=0}^{2} |B^{(z)}(p,q)|.
\label{RD} 
\end{equation}
\subsection{Range-Doppler Estimation}
To obtain the range, Doppler and AoA information of the hand in every measurement-cycle, we select $K$ points from $RD(p, q)$, which have the largest magnitudes. The parameter $K$ is predefined, and its choice will be discussed in Section \ref{derK}. Then, we extract the range, Doppler frequencies and the magnitudes of those $K$ points, which are denoted as $\hat{f}_{\mathrm{r}k}$, $\hat{f}_{\mathrm{D}k}$ and $A_k$, respectively, where $k = 1, \cdots, K$.

\subsection{Azimuth and Elevation Angle Estimation}
The AoA can be calculated from the phase difference of extracted points in the same positions of complex-valued RD spectrums belonging to two receive antennas. The AoA in azimuth and elevation of the $k$-th point can be calculated as:
\begin{equation}
\hat{\phi}_k= \arcsin\left( \frac{\left( \psi\left( a_k^{(1)}\right) -\psi\left( a_k^{(2)}\right) \right) \lambda}{2 \pi d} \right),
\label{azi} 
\end{equation}
\begin{equation}
\hat{\theta}_k= \arcsin\left( \frac{\left( \psi\left( a_k^{(1)}\right) -\psi\left( a_k^{(0)}\right) \right) \lambda}{2 \pi d} \right),
\label{ele} 
\end{equation}
respectively, where $\psi (\cdot)$ stands for the phase of a complex value, $a_k^{(z)}$ is the complex amplitude $B^{(z)}\left( \hat{f}_{\mathrm{r}k}, \hat{f}_{\mathrm{D}k}\right)$ from the $z$-th receive antenna.
\subsection{Feature Cube}
\label{encoder}
\begin{figure}[htbp]
	\centering
	\includegraphics[width=8.5cm]{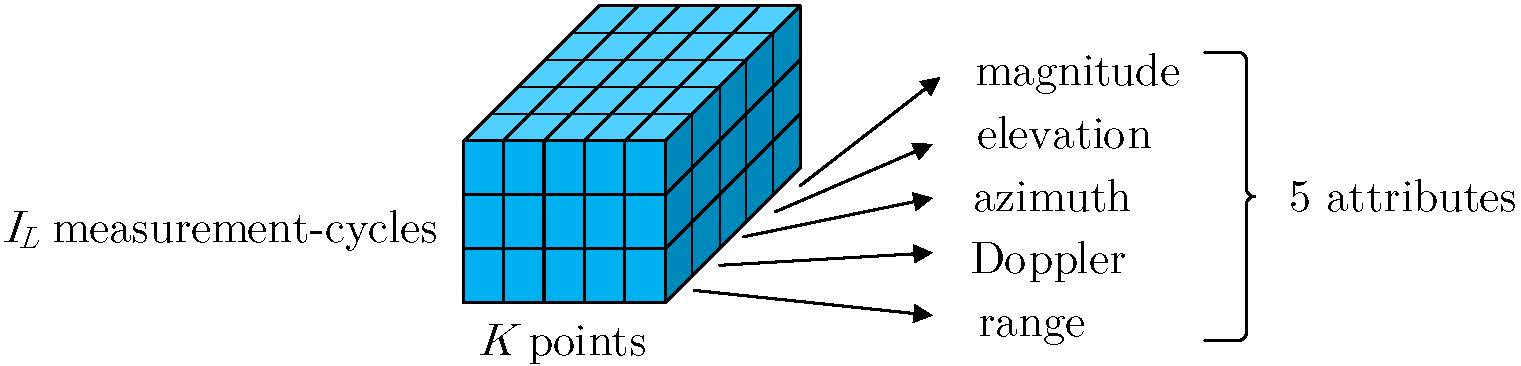}
	\caption{Structure of feature cube $\mathcal{V}$.}
	\label{featurecube}
\end{figure}

As a consequence, in every measurement-cycle, the $k$-th point in $RD(p, q)$ has five attributes, i.e., range, Doppler, azimuth, elevation and magnitude. As depicted in Fig.~\ref{featurecube}, we encode the range, Doppler, azimuth, elevation and magnitude of those $K$ points with the largest magnitudes in $RD(p,q)$ along $I_L$ measurement-cycles into the feature cube $\mathcal{V}$ with dimension $I_L \times K \times 5$. The $\mathcal{V}$ has five channels corresponding to five attributes and each element in $\mathcal{V}$ at the $l$-th measurement-cycle can be described as:
\begin{equation}
\begin{split}
\mathcal{V}(l,k,1) &= \hat{f}_{\mathrm{r}k}, \quad \mathcal{V}(l,k,2) = \hat{f}_{\mathrm{D}k}, \quad \mathcal{V}(l,k,3) = \hat{\phi}_{k},\\
\mathcal{V}(l,k,4) &= \hat{\theta}_{k}, \quad \mathcal{V}(l,k,5) = A_{k},\\
\end{split}
\end{equation}
where $l =1, \cdots, I_L$.
\section{Hand Activity Detection}
\label{detection}
\subsection{Problem Statement}
\begin{figure}[t]
	\captionsetup[subfloat]{labelformat=parens,farskip=0pt,nearskip=0pt}
	\setlength{\belowcaptionskip}{0pt}
	\centering
	\subfloat[]{\includegraphics[width=0.5\columnwidth]{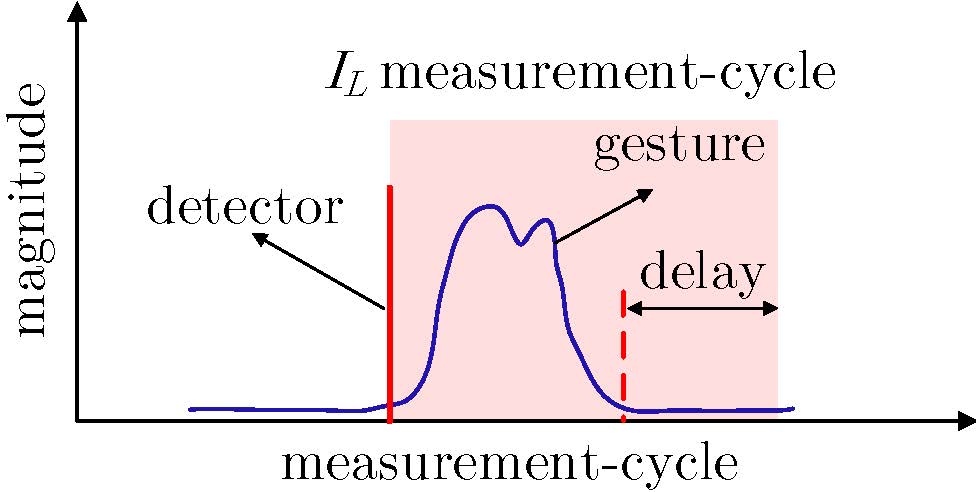}\label{wrong}}
	\hfil
	\subfloat[]{\includegraphics[width=0.5\columnwidth]{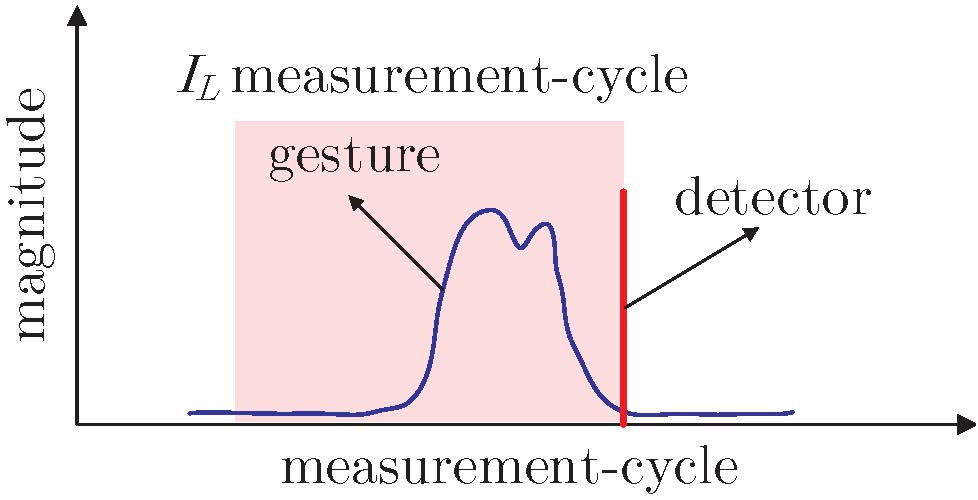}\label{right}}	
	\caption{(a) A delay occurs before forwarding the gesture data to the classifier when we detect the start time-stamp of the gesture. (b) The gesture data is directly forwarded to the classifier without delay when we detect the tail of the gesture.}
	\label{vad1}
\end{figure}

Similar to voice activity detection in the automatic speech recognition system, our gesture recognition system also needs to detect some hand activities in advance, before forwarding the data to the classifier. It helps to design a power-efficient gesture recognition system, since the classifier is only activated when a gesture is detected rather than keeping it active for every measurement-cycle. The state-of-the-art event detection algorithms usually detect the start time-stamp of an event. For example, the authors in \cite{vaezi2015comparison} used the STA/LTA and power spectral density methods to detect when a micro-seismic event occurs. In the case of radar-based gesture recognition, we could also theoretically detect the start time-stamp of a gesture and consider that a gesture event occurs within the following $I_L$ measurement-cycles. However, detecting the start-stamp and forwarding the hand data in the following $I_L$ measurement-cycles to the classifier could cause a certain time delay, since the time duration of designed gestures is usually different. As illustrated in Fig.~\ref{vad1}\subref{wrong}, due to the facts that the proposed multi-feature encoder requires $I_L$ measurement-cycles and the duration of the gesture is usually shorter than $I_L$, a delay occurs, if we detect the start time-stamp of the gesture. Therefore, as depicted in Fig.~\ref{vad1}\subref{right}, to reduce the time delay, our proposed HAD algorithm is designed to detect when a gesture finishes, i.e., the tail of a gesture, rather than detecting the start time-stamp.
\subsection{STA/LTA-based Gesture Detector}
\begin{figure}[htbp]
	\centering
	\includegraphics[width=9.5cm]{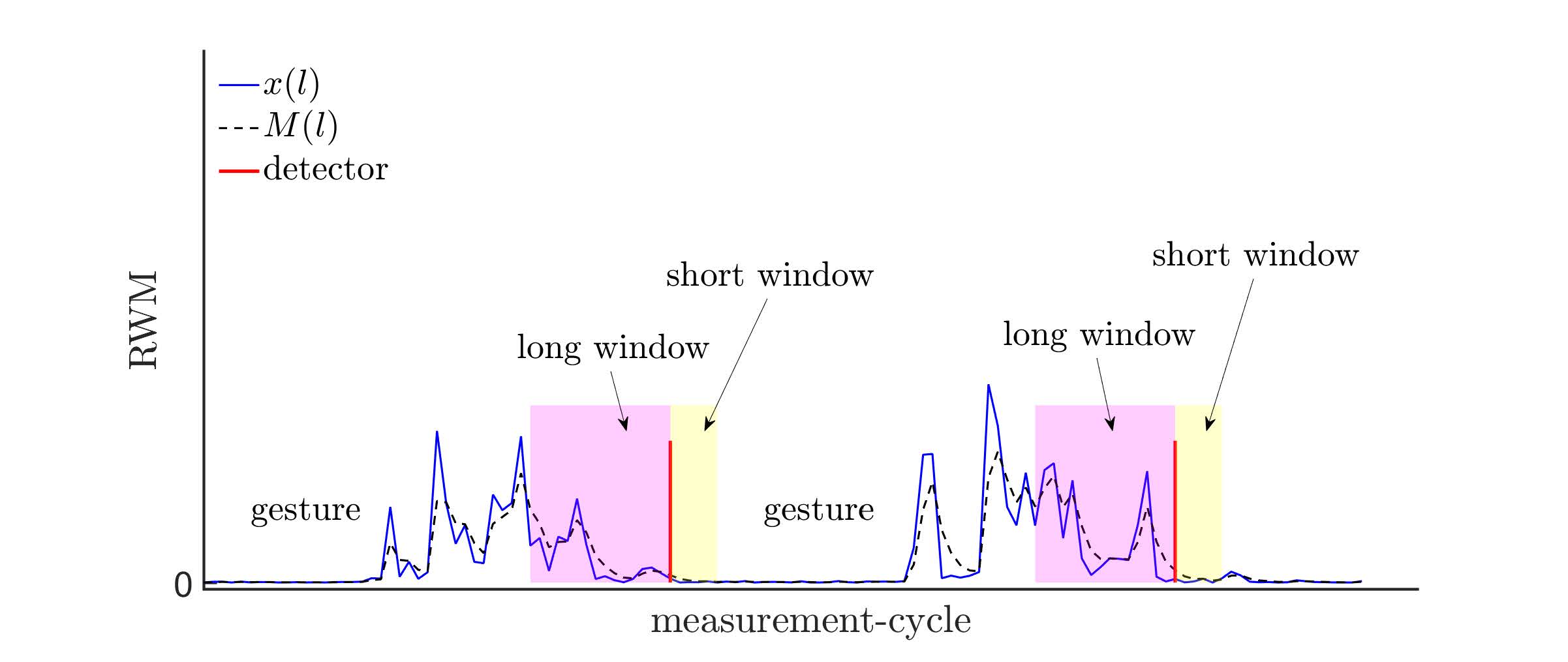}
	\caption{The tails of two gestures are detected via the proposed gesture detector.}
	\label{vad}
\end{figure}

We propose a STA/LTA-based gesture detector to detect the tail of a gesture. The exponential moving average (EMA) is used to detect the change of the magnitude signal at the $l$-th measurement-cycle, which is given as: 
\begin{equation}
M(l) = (1- \alpha) M(l-1) + \alpha x(l),
\end{equation}
where $\alpha \in \left[ 0, 1\right] $ is the predefined smoothing factor, $x(l)$ is the range-weighted magnitude (RWM), and it is defined as:
\begin{equation}
	x(l) = A_\mathrm{max} f^\beta_{\mathrm{r}_\mathrm{max}}, \quad A_\mathrm{max} = \max_k \left\lbrace A_k\right\rbrace,
	\label{rwa}
\end{equation}
where $A_\mathrm{max}$ represents the maximal magnitude among $K$ points in $RD(p,q)$ at $l$-th measurement-cycle, $f_{\mathrm{r}_\mathrm{max}}$ denotes the range corresponding to $A_\mathrm{max}$, and the predefined coefficient $\beta$ denotes the compensation factor. The radar cross section (RCS) of a target is independent of the propagation path loss between the radar and the target. According to the radar equation \cite{skolnik1990radar}, the measured magnitude of a target is a function of many arguments, such as the path loss, RCS, etc. As deduced in \eqref{rwa}, we have built a coarse estimate of the RCS by multiplying the maximal range information with its measured magnitude to partially compensate the path loss. Furthermore, we define the $\text{STA}(l)$ and $\text{LTA}(l)$ as the mean EMA in short and long windows at the $l$-th measurement-cycle:
\begin{equation}
\text{STA}(l) = \frac{1}{L_1}\sum_{ll=l+1}^{l+L_1}M(ll),\quad \text{LTA}(l) = \frac{1}{L_2}\sum_{ll = l-L_2+1}^{l}M(ll),
\end{equation}
respectively, where $L_1$ and $L_2$ are the length of the short and long window. The tail of a gesture is detected, when the following conditions are fulfilled:
\begin{equation}
\sum_{ll = l - L_2+1}^{l} x(ll) \geq \gamma_1 \quad \text{and} \quad  \frac{\text{STA}(l)}{\text{LTA}(l)} \leq \gamma_2, 
\label{stalta}
\end{equation}
where $\gamma_1$ and $\gamma_2$ are the predefined detection thresholds. Fig.~\ref{vad} illustrates that the tails of two gestures are detected via the proposed STA/LTA gesture detector. According to \eqref{stalta}, one condition of detecting the tail of a gesture is that, the average of RWM in the long window exceeds the threshold $\gamma_1$. It means that a hand motion appears in the long window. The other condition is that, the ratio of the mean EMA in the short window and that in the long window is lower than the threshold $\gamma_2$. In other words, it detects when the hand movement finishes. In practice, the parameters $\beta$, $\gamma_1$ and $\gamma_2$ in our HAD algorithm should be thoroughly chosen according to different application scenarios.

\section{Supervised Learning}
\label{supervised}
\begin{figure*}[htbp]
	\centering
	\includegraphics[width=17.5cm]{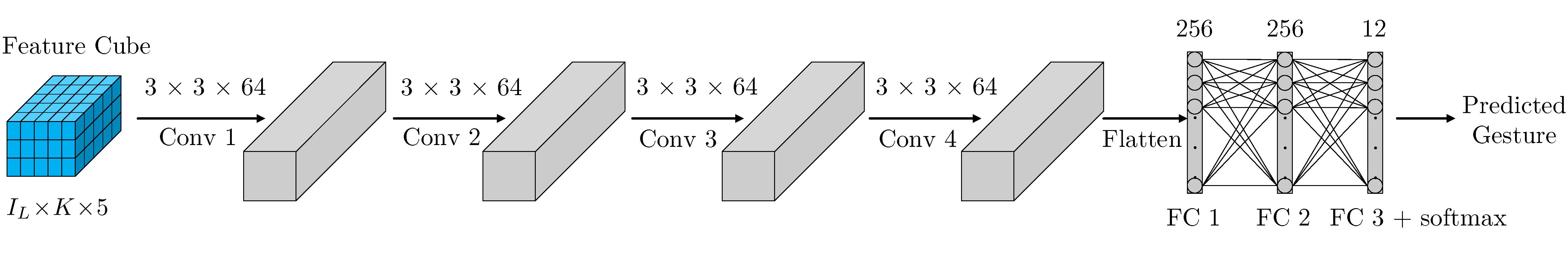}
	\caption{Structure of the shallow CNN taking the feature cube as input.}
	\label{2dcnn}
\end{figure*}

As discussed in Section \ref{encoder}, the feature cube obtained by the multi-feature encoder has a dimension of $I_L \times K \times 5$. Thus, we could simply use the CNN for classification without any reshaping operation. The structure of the CNN can be seen in Fig.~\ref{2dcnn}. We employ four convolutional (Conv) layers, each of that has a kernel size $3 \times 3$ and the number of kernels in each Conv layer is 64. In addition, the depth of the first kernel is five, since the input feature cube has five channels (i.e., range, Doppler, azimuth, elevation and magnitude), while that of the other kernels in the following three Conv layers is 64. We choose the rectified linear unit (ReLU) \cite{nair2010rectified} as activation function, since it solves the problem of gradient vanishing and is able to accelerate the convergence speed of training \cite{xu2015empirical}. Then, the last Conv layer is connected by two fully-connected (FC) layers, either of which has 256 hidden units and is followed by a dropout layer for preventing the network from overfitting. The third FC layer with a softmax function is utilized as the output layer. The number of hidden units in the third FC layer is designed to be in accordance with the number of classes in the dataset. The softmax function normalizes the output of the last FC layer to a probability distribution over the classes.

Through thoroughly network tuning (e.g., number of hidden layers, number of hidden units, depth number), we construct the CNN structure as shown in Fig.~\ref{2dcnn}. The designed network should (a) take the feature cube as input, (b) achieve a high classification accuracy, (c) consume few computational resources, and (d) be deployable in the edge-computing platform. In Section~\ref{experiment}, we will show that the designed network in Fig.~\ref{2dcnn} fulfills these criteria.
\section{Scenario and Gesture Dataset Description}
\label{dataset}
\begin{figure}[h]
	\centering
	\includegraphics[width=1.5in]{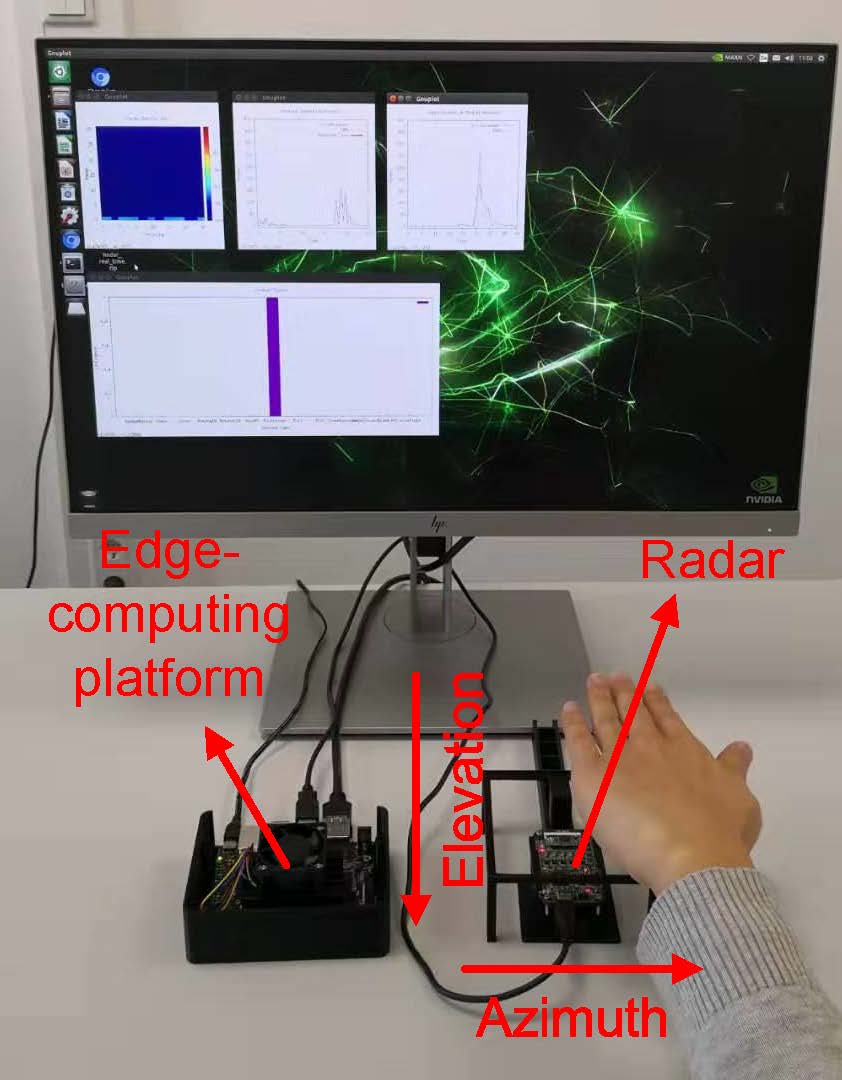}
	\caption{Experiment scenario of the radar-based gesture recognition system.}
	\label{mea}
\end{figure}

\begin{figure}[h]
	\centering
	\subfloat[]{\includegraphics[width=0.22\columnwidth]{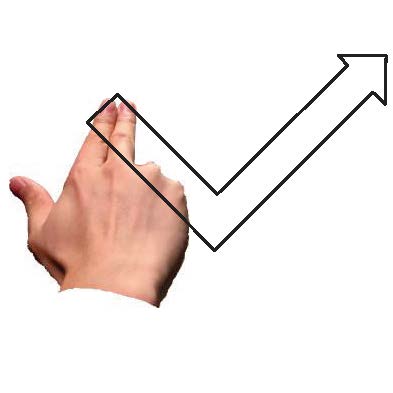}}\hspace{0.2em}
	\subfloat[]{\includegraphics[width=0.22\columnwidth]{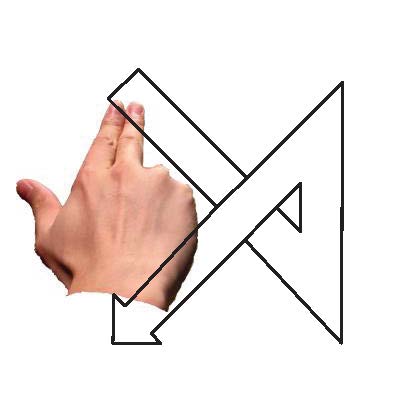}}\hspace{0.2em}
	\subfloat[]{\includegraphics[width=0.22\columnwidth]{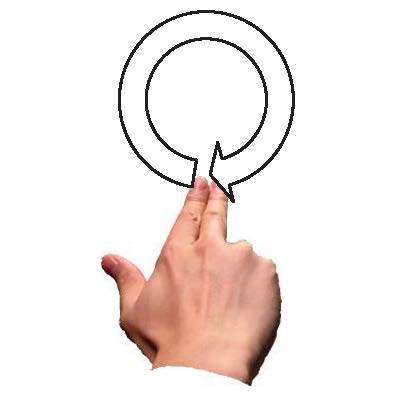}}\hspace{0.2em}
	\subfloat[]{\includegraphics[width=0.22\columnwidth]{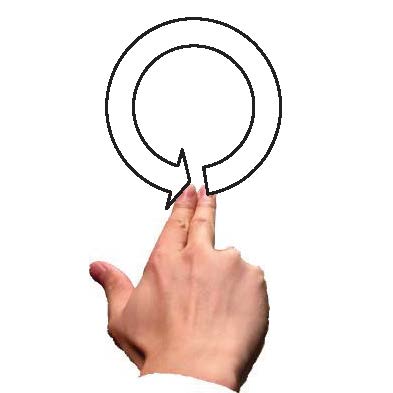}}\hspace{0.2em}	%\hfill
	\subfloat[]{\includegraphics[width=0.22\columnwidth]{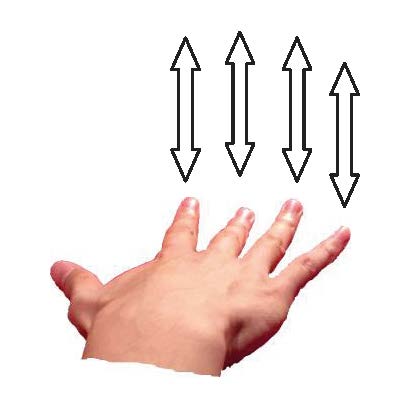}}\hspace{0.2em}	
	\subfloat[]{\includegraphics[width=0.22\columnwidth]{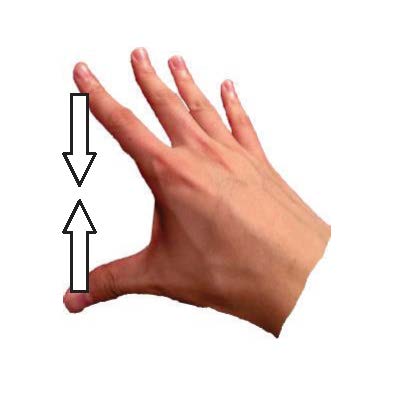}}\hspace{0.2em}
	\subfloat[]{\includegraphics[width=0.22\columnwidth]{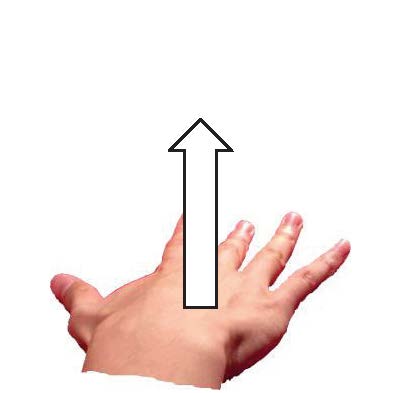}}\hspace{0.2em}	
	\subfloat[]{\includegraphics[width=0.22\columnwidth]{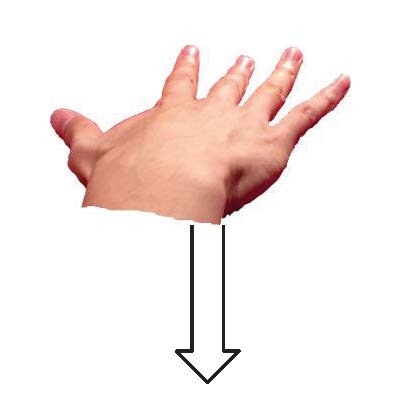}}	\hspace{0.2em}%\hfill
	\subfloat[]{\includegraphics[width=0.22\columnwidth]{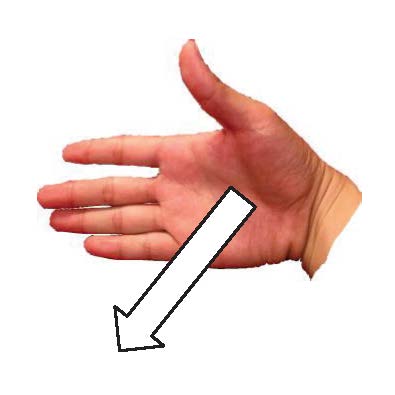}}\hspace{0.2em}
	\subfloat[]{\includegraphics[width=0.22\columnwidth]{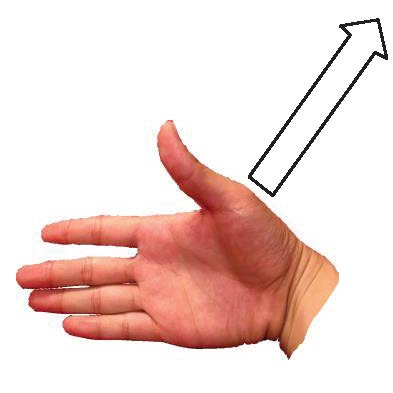}}	\hspace{0.2em}
	\subfloat[]{\includegraphics[width=0.22\columnwidth]{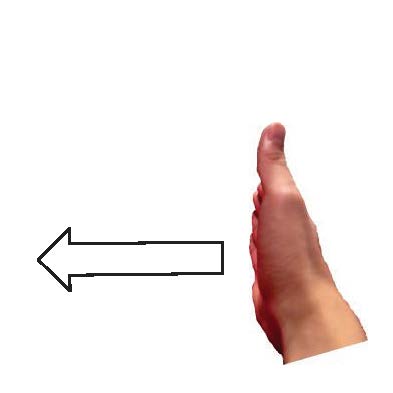}}\hspace{0.2em}	
	\subfloat[]{\includegraphics[width=0.22\columnwidth]{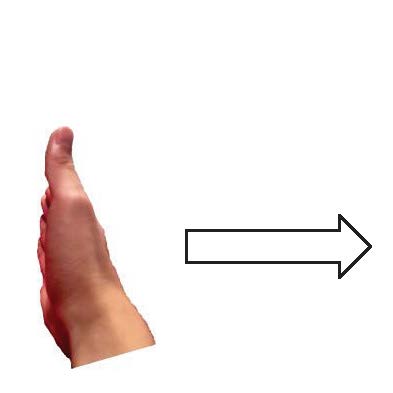}}
	\caption{(a) Check. (b) Cross. (c) Rotate CW. (d) Rotate CCW. (e) Moving fingers. (f) Pinch index. (g) Pull. (h) Push. (i) Swipe BW. (j) Swipe FW. (k) Swipe LT. (l) Swipe RT.}
	\label{ges_mea}
\end{figure}

As illustrated in Fig.~\ref{mea}, we used the 60 GHz FMCW radar in Fig.~\ref{evalkit} to recognize gestures. Our radar system has a detection range up to 0.9 m and an approx. $120^{\circ}$ antenna beam width in both azimuth and elevation directions. The parameter setting used in the waveform design is presented in Table \ref{radar parameter}, where the pulse repetition interval (PRI) is 34 ms.
\begin{table}[h]
	\centering
	\caption{Transmit signal design and radar parameters}
	\setlength{\tabcolsep}{0.8em}
	\begin{tabular}[t]{lccccccccc}
		\toprule
		Transmit & $f_\mathrm{c}$&$f_\mathrm{B}$&$T_\mathrm{c} $&$I_\mathrm{s}$&$I_\mathrm{c}$&$\mathrm{PRI}$\\
		signal design&60 GHz&5 GHz& 432 \textmu s & 32 & 32 & 34 ms\\
		\multirow{2}{*}{Resolution}& $	\Delta r$&$\Delta v_\mathrm{r}$\\
		& 3 cm &  18 cm/s\\	
		\bottomrule
	\end{tabular}
	\label{radar parameter}
\end{table}
The radar is connected with an edge-computing platform, i.e., NVIDIA Jetson Nano, which is equipped with Quad-core ARM A57 at 1.43 GHz as central processing unit (CPU), 128-core Maxwell as graphics processing unit (GPU) and 4 GB memory. We have built our entire radar-based gesture recognition framework described in Fig.~\ref{framework} in the edge-computing platform in C/C++. The proposed multi-feature encoder and HAD have been implemented in a straightforward manner without any runtime optimization, while the implementation of the CNN is supported by TensorRT developed by NVIDIA. In addition, as depicted in Fig.~\ref{ges_mea}, we designed 12 gestures, which are (a) Check, (b) Cross, (c) Rotate clockwise (CW), (d) Rotate counterclockwise (CCW), (e) Moving fingers (FG), (f) Pinch index, (g) Pull, (h) Push, (i) Swipe backward (BW), (j) Swipe forward (FW), (k) Swipe left (LT) and (l) Swipe right (RT). We invited 20 human subjects including both genders with various heights and ages to perform these gestures. Among 20 subjects, the ages range from 20 to 35 years old, and the heights are from 160 cm to 200 cm. We divided the 20 subjects into two groups. In the first group, ten subjects were taught how to perform gestures in a normative way. Whereas, in the second group, in order to increase the diversity of the dataset, only an example for each gesture was demonstrated to the other ten subjects and they performed gestures using their own interpretations. Self-evidently, their gestures were no longer as normative as the ones performed by the ten taught subjects. Furthermore, every subject repeated each gesture 30 times. Therefore, the total number of realizations in our gesture dataset is (12 gestures)$\times$(20 people)$\times$(30 times), namely 7200. We also found out that the gestures performed in our dataset take less than 1.2 s. Thus, to ensure that the entire hand movement of a gesture is included in the observation time, we set $I_L$ to 40, which amounts to a duration of 1.36 s (40 measurement-cycles $\times$ 34 ms).

\section{Experimental Results}
\label{experiment}
In this section, the proposed approach is evaluated regarding a twofold objective: first, its performance is thoroughly compared with benchmarks in literature through an off-line cross-validation, and secondly, its real-time capability is investigated with an on-line performance test. In Section \ref{derK}, we discuss how the parameter $K$ affects the classification accuracy. In Section \ref{off-line}, we compare our proposed algorithm with the state-of-the-art radar-based gesture recognition algorithms in terms of classification accuracy and computational complexity based on leave-one-out cross-validation (LOOCV). It means that, in each fold, we use the gestures from one subject as test set, and the rest as training set. In addition, Section \ref{online} describes the real-time evaluation results of our system. The performances of taught and untaught subjects are evaluated separately. We randomly selected eight taught and eight untaught subjects as training sets, while the remaining two taught and two untaught subjects are test sets. In real-time performance evaluation, we performed the hardware-in-the-loop (HIL) test, and fed the raw data recorded by the radar from the four test subjects into our edge-computing platform.
\subsection{Determination of Parameter $K$}
\label{derK}
\begin{figure}[h]
	\centering
	\includegraphics[width=3.4in]{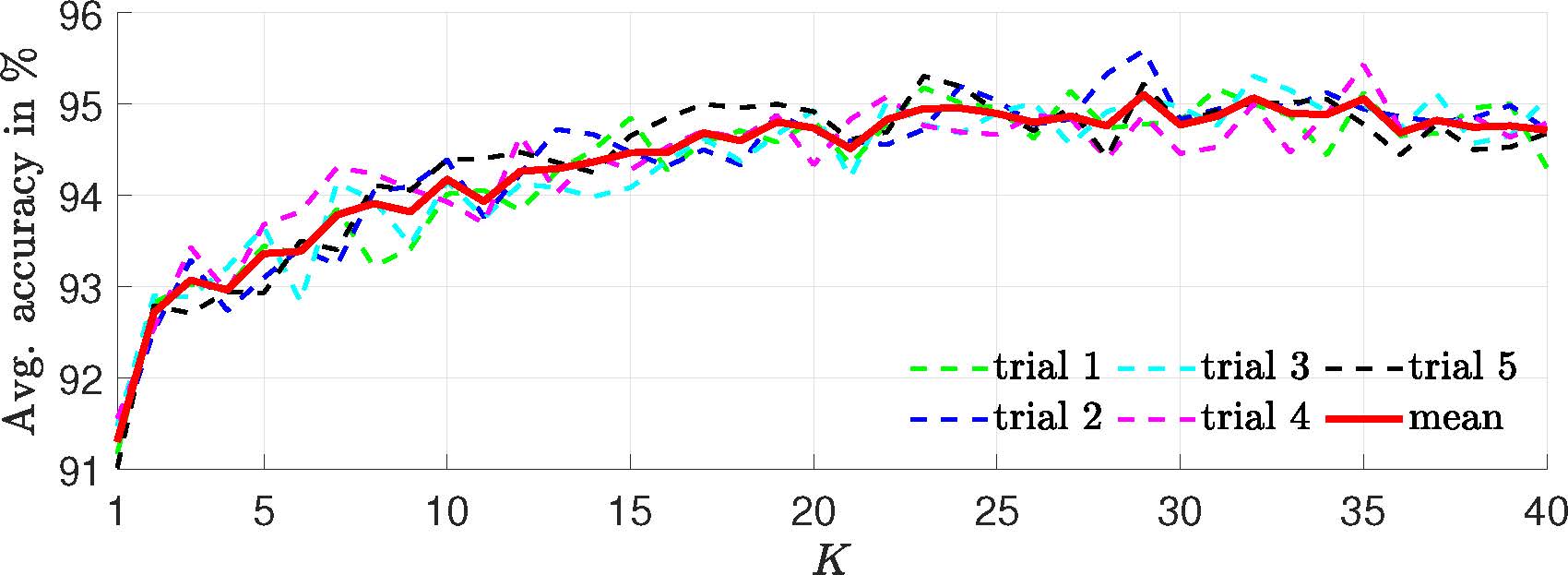}
	\caption{Determination of the number of extracted points $K$ from $RD(p, q)$.}
	\label{park}
\end{figure}

As described in Section \ref{preprocessing}, we extract $K$ points with the largest magnitudes from $RD(p,q)$, to represent the hand information in a single measurement-cycle. We define the average (avg.) accuracy as the avg. classification accuracy across the 12 gestures based on LOOCV. In Fig.~\ref{park}, we let $K$ vary from 1 to 40, and compute the avg. accuracy in five trials. It can be seen that the mean avg. accuracy over five trials keeps increasing and reaches approx. 95\%, when $K$ is 25. After that, increasing $K$ can barely improve the classification accuracy. As a result, in order to keep low computational complexity of the system and achieve a high classification accuracy, we set $K$ to 25. It results that the feature cube $\mathcal{V}$ in our proposed multi-feature encoder has a dimension of $40 \times 25 \times 5$.

\subsection{Off-line Performance Evaluation}
\label{off-line}
In the off-line case, we assumed that each gesture is perfectly detected by the HAD algorithm and compared our proposed multi-feature encoder + CNN with the 2-D CNN + LSTM \cite{wang2016interacting}, the 3-D CNN + LSTM \cite{zhang2018latern}, 3-D CNN + LSTM (with AoA) and shallow 3-D CNN + LSTM (with AoA) in terms of the avg. classification accuracy and computational complexity based on LOOCV. In our proposed multi-feature encoder + CNN, the feature cube $\mathcal{V}$, which has the dimension of $40 \times 25 \times 5$, was fed into the CNN described in Fig.~\ref{2dcnn}. The input of the 2-D CNN + LSTM \cite{wang2016interacting} and the 3-D CNN + LSTM \cite{zhang2018latern} is the RD spectrum sequence over 40 measurement-cycles, which has the dimension of $40 \times 32 \times 32 \times 1$. Since \cite{zhang2018latern} did not include any AoA information in their system for gesture classification, the comparison might not be fair. Thus, we added the AoA information according to \eqref{azi} and \eqref{ele} to the RD spectrum sequence. It results in the input of the 3-D CNN + LSTM (with AoA) and shallow 3-D CNN + LSTM (with AoA) having the dimension of $40 \times 32 \times 32 \times 3$, where the second and the third channel contain the AoA information in azimuth and elevation dimension, respectively. The shallow 3-D CNN + LSTM (with AoA) is designed to having comparable computational complexity as that of the proposed multi-feature encoder + CNN but with reduced classification accuracy. To achieve a fair comparison, we optimized the structures and the hyper-parameters as well as the training parameters of those models. The CNN demonstrated in Fig.~\ref{2dcnn} in the proposed approach was trained for 15000 steps based on the back propagation \cite{lecun1989backpropagation} using the Adam optimizer \cite{kingma2014adam} with an initial learning rate of $1\times 10^{-4}$, which degraded to $10^{-5}$, $10^{-6}$ and $10^{-7}$ after 5000, 8000 and 11000 steps, respectively. The batch size is 128.

\subsubsection{Classification Accuracy and Training Loss Curve}
In Table \ref{avgacc}, we present the classification accuracy of each type of gesture based on the algorithms mentioned above. The avg. accuracies of the 2-D CNN + LSTM \cite{wang2016interacting} and 3-D CNN + LSTM \cite{zhang2018latern} are only $78.50\%$ and $79.76\%$, respectively. Since no AoA information is utilized, the Rotate CW and Rotate CCW can hardly be distinguished, and similarly the four Swipe gestures can hardly be separated, either. On the contrary, considering the AoA information, the multi-feature encoder + CNN, the 3-D CNN + LSTM (with AoA) and the shallow 3-D CNN + LSTM (with AoA) are able to separate the two Rotate gestures, and the four Swipe gestures. It needs to be mentioned that the avg. accuracy of our proposed multi-feature encoder is almost the same as that of the 3-D CNN + LSTM with (AoA). However, it will be shown in the following section that our approach requires much less computational resources and memory than those of the other approaches. 

\begin{figure}[h]
	\centering
	\includegraphics[width=3.4in]{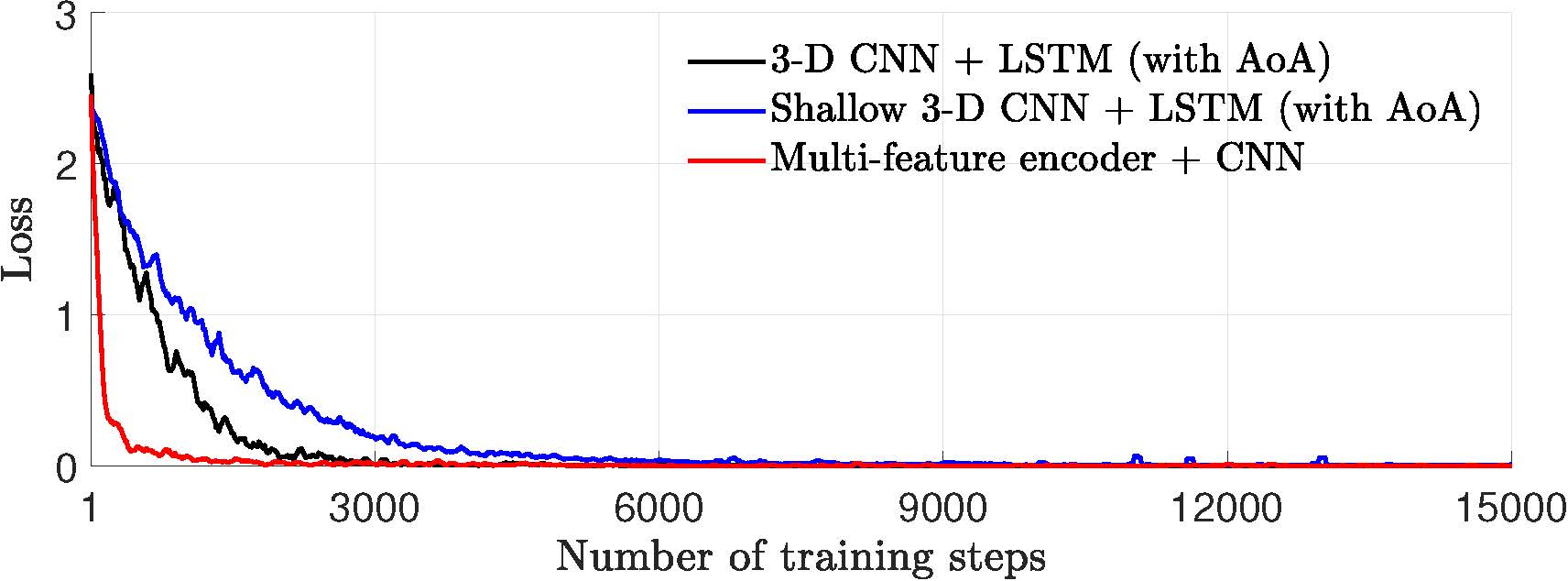}
	\caption{Comparison of training loss curves.}
	\label{trainingcurve}
\end{figure}

What's more, in Fig.~\ref{trainingcurve}, we plot the training loss curves of the three structures of neural networks. It can be seen that the loss of the proposed CNN in Fig.~\ref{2dcnn} has the fastest rate of convergence among the three structures of neural networks and approaches to zero at around the $2000$-th training step. Unlike the input of the 3-D CNN + LSTM (with AoA) and shallow 3-D CNN + LSTM (with AoA), the feature cube contains sufficient gesture characteristics in spite of its compact form ($40 \times 25 \times 5$). It results that the CNN in Fig.~\ref{2dcnn} is easier to be trained than the other neural networks, and it achieves a high classification accuracy.

\begin{table*}[h]
	\centering	
	\caption{Classification accuracy in $\%$ of each gesture obtained by different gesture recognition frameworks}
	\setlength{\tabcolsep}{0.35em}
	\begin{tabular}[t]{lccccccccccccc}
		\toprule
		&avg. acc.&(a)&(b)&(c)&(d)&(e)&(f)&(g)&(h)&(i)&(j)&(k)&(l)\\
		\midrule
		2-D CNN + LSTM \cite{wang2016interacting} &78.50&85.17&82.67&60.67&55.50&93.33&95.00&90.67&91.17&66.83&75.33&67.50&78.17\\
		%CNN + LSTM (with AoA) &79.76&83.17&87.17&62.83&57.17&93.50&97.17&93.17&92.17&63.67&77.33&69.33&80.50\%\\		
		3-D CNN + LSTM \cite{zhang2018latern} &79.76&83.17&87.17&62.83&57.17&93.50&97.17&93.17&92.17&63.67&77.33&69.33&80.50\\
		3-D CNN + LSTM (with AoA) &95.57&96.67&95.17&97.17&96.33&92.00&95.17&94.67&95.17&94.50&93.50&98.50&98.0\\
		Shallow 3-D CNN + LSTM (with AoA)&94.36&95.33&92.0&96.67&96.83&93.33&94.83&95.83&93.0&90.0&89.17&98.83&96.50\\
		Multi-feature encoder + CNN &95.79&96.50&97.83&95.83&96.83&96.17&95.50&93.17&96.50&92.67&92.67&98.17&97.67\\
		\bottomrule
	\end{tabular}
	\label{avgacc}
\end{table*}
\begin{table*}[h]
	\centering
	\caption{Different neural network structures for radar-based gesture recognition}
	\setlength{\tabcolsep}{0.8em}
	\begin{tabular}[t]{lccccccccccccc}
		\toprule
		Layers &3-D CNN + LSTM (with AoA)&Shallow 3-D CNN + LSTM (with AoA)&Multi-feature encoder + CNN\\
		\midrule
		input &$40\times32\times32\times3$&$40\times32\times32\times3$& $40\times 25 \times 5$\\
		1 &3-D Conv1 3 $\times$ 3 $\times$ 3 $\times$ 16&3-D Conv1 3 $\times$ 3 $\times$ 3 $\times$ 16&Conv1 3 $\times$ 3 $\times$ 64\\ 
		2 &3-D Conv2 3 $\times$ 3 $\times$ 3 $\times$ 32&3-D Conv2 3 $\times$ 3 $\times$ 3 $\times$ 32&Conv2 3 $\times$ 3 $\times$ 64\\
		3 &3-D Conv3 3 $\times$ 3 $\times$ 3 $\times$ 64&3-D Conv3 3 $\times$ 3 $\times$ 3 $\times$ 64&Conv3 3 $\times$ 3 $\times$ 64\\
		4 &FC1 512&FC1 64&Conv4 3 $\times$ 3 $\times$ 64\\
		5 &FC2 512&LSTM 32&FC1 256\\
		6 &LSTM 512&FC2 12 - softmax&FC2 256\\
		7 &FC3 12 - softmax&-&FC3 12 - softmax\\
		\midrule
		%Computational complexity &\\
		GFLOPs & 2.89 & 0.34 & 0.26\\
		Size & 109 MB & 101 MB& 4.18 MB\\
		\bottomrule
	\end{tabular}
	\label{structure}
\end{table*}
\subsubsection{Confusion Matrix}
\begin{figure*}[t]
	\captionsetup[subfloat]{labelformat=parens,farskip=0pt,nearskip=0pt}
	\setlength{\belowcaptionskip}{0pt}
	\centering
	\subfloat[]{\includegraphics[width=1\columnwidth]{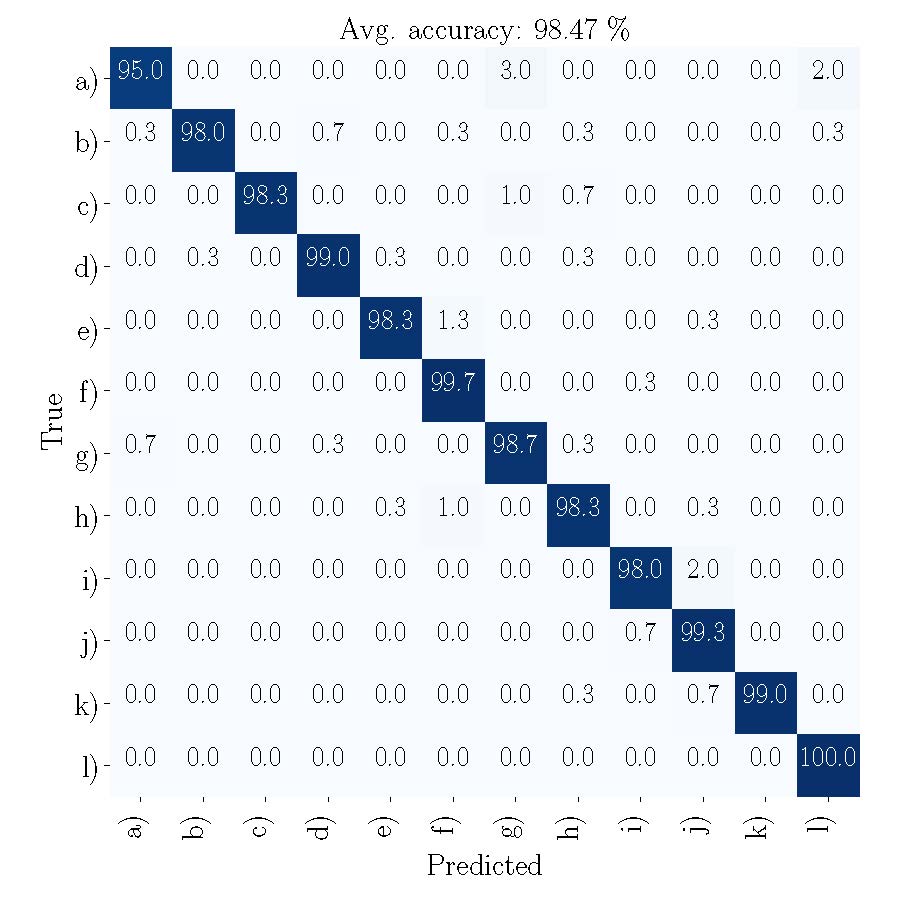}\label{taught}}
	\hfil
	\subfloat[]{\includegraphics[width=1\columnwidth]{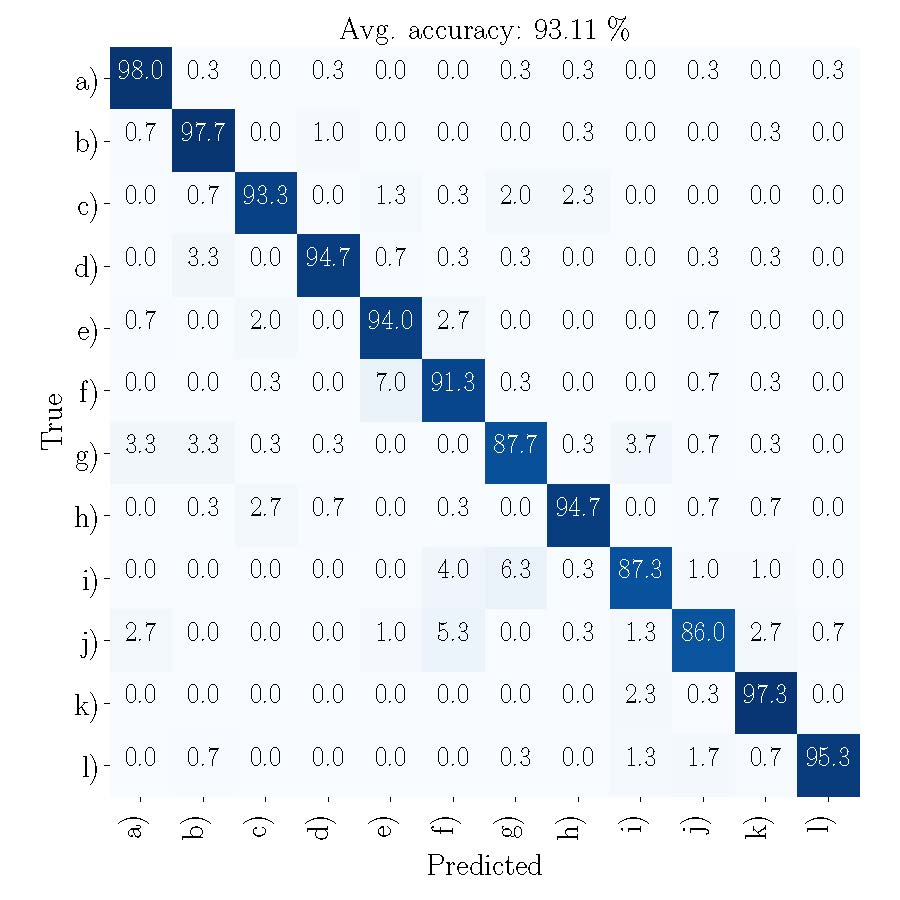}\label{untaught}}
	\caption{Confusion matrices obtained by the proposed multi-feature encoder + CNN based on LOOCV. (a) 10 taught subjects. (b) 10 untaught subjects.}
	\label{confusion_matrix}
\end{figure*}

In Fig.~\ref{confusion_matrix}, we plotted two confusion matrices for ten taught and ten untaught subjects based on our proposed multi-feature encoder + CNN. It could be observed that, for the normative gestures performed by the ten taught subjects, we could reach approx. 98.47\% avg. accuracy. Although we could observe an approx. 5\% degradation in avg. accuracy in Fig.~\ref{confusion_matrix}\subref{untaught}, where the gestures to be classified are performed by ten untaught subjects, it still has 93.11\% avg. accuracy.
\subsubsection{Computational Complexity and Memory}
The structures of the 3-D CNN + LSTM (with AoA), shallow 3-D CNN + LSTM (with AoA) and the proposed multi-feature encoder + CNN are presented in Table \ref{structure}. We evaluated their computational complexity and required memory in line with the giga floating point operations per second (GFLOPs) and the model size. The GFLOPs of different models were calculated by the built-in function in TensorFlow, the model size is observed through TensorBoard \cite{abadi2016tensorflow}. Although the 3-D CNN + LSTM (with AoA) offers almost the same classification accuracy as that of the proposed multi-feature encoder + CNN, it needs much more GFLOPs than that of the multi-feature encoder + CNN ($2.89$ GFLOPs vs. $0.26$ GFLOPs). Its model size is also much larger than that of the proposed approach (109 MB vs. 4.18 MB). Although we could reduce its GFLOPs using a shallow network structure, such as the shallow 3-D CNN + LSTM (with AoA) in Table \ref{structure}, it results in the degradation of classification accuracy (94.36\%), as can be seen in Table \ref{avgacc}. We also found out that the CNN used in our approach has the least model size, since its input dimension is much smaller than that of other approaches. On the contrary, the input of the 3-D CNN + LSTM (with AoA) contains lots of zeros due to the sparsity of RD spectrums. Such large volumes usually need large amounts of coefficients in neural networks. Whereas, we exploit the hand information in every measurement-cycle using only 25 points, and the input dimension of the CNN is only $40 \times 25 \times 5$, which requires much less computational complexity than the other approaches.  

\subsection{Real-time Performance Evaluation}
\label{online}
As mentioned above, subjects are divided into taught and untaught groups, and each has ten subjects. In each group, eight subjects are randomly selected as training set, and the remaining two subjects constitute the test set, resulting in either group having 720 true gestures in the test set. In the HIL context, we directly fed the recorded raw data from the four test subjects into the edge-computing platform. In the real-time case, the system should be robust enough to distinguish true gestures from random motions (RMs). Thus, we also included a certain amount of RMs as negative samples during the training phase. The scale of RMs and true gestures is around 1:3.  
\subsubsection{Precision, Recall and $F_1$-score}
To quantitatively analyze the real-time performance of our system, we introduce the precision, recall and $F_1$-score, which are calculated as: 
\begin{equation}
\begin{split}
\text{precision} &= \frac{\mathrm{TP}}{\mathrm{TP} + \mathrm{FP}}, \quad   \text{recall} =  \frac{\mathrm{TP}}{\mathrm{TP} + \mathrm{FN}},\\
&F_1\text{-score} = 2 \cdot \frac{\text{precision} \cdot \text{recall}}{\text{precision} + \text{recall}},
\end{split}
\end{equation}
where $\mathrm{TP}$, $\mathrm{FP}$ and $\mathrm{FN}$ denote the number of true positive, false positive, and false negative estimates. For two subjects in the test set, we have 60 realizations for each gesture. It means that $\mathrm{TP}+\mathrm{FN} = 60$. As presented in Table \ref{f1score}, the avg. precision and recall over 12 types of gestures using two taught subjects as test set are 93.90\% and 94.44\%, respectively, while those using two untaught subjects as test set are 91.20\% and 86.11\%. It needs to be mentioned that, the off-line avg. accuracies in Fig.~\ref{confusion_matrix}, namely 98.47 \% and 93.11\%, can also be regarded as the recall in taught and untaught cases. After comparing with the recall in the off-line case, we could observe an approx. 4\%  and 7\% degradation in recall in the real-time case considering both the taught and untaught subjects. The reason is that, in the off-line performance evaluation, we assumed that each gesture is detected perfectly. However, in the real-time case, the recall reduction is caused by the facts that our HAD performance miss-detected some gestures or incorrectly triggered the classifier even when the gesture was not completely finished. For example, due to the small movement of the hand, the HAD sometimes failed to detect the gesture "Pinch index". Similarly, the recall of the gesture "Cross" is also impaired, since the gesture "Cross" has a turning point, which leads to a short pause. In some cases where the subject performs the gesture "Cross" with low-velocity, the HAD would incorrectly consider the turning point as the end of "Cross", resulting in a wrong classification. Overall, in both taught and untaught cases, the $F_1$-score of our radar-based gesture recognition system reaches 94.17\% and 88.58\%, respectively.  
\begin{table*}[!htbp]
	\centering	
	\caption{Precision, recall and $F_1$-score in $\%$ of the real-time radar-based gesture recognition system in both taught and untaught cases}
	\setlength{\tabcolsep}{0.35em}
	\begin{tabular}[t]{lccccccccccccccc}
		\toprule
		&&avg.&(a)&(b)&(c)&(d)&(e)&(f)&(g)&(h)&(i)&(j)&(k)&(l)&$F_1$-score\\
		\midrule
		\multirow{2}{*}{Taught} &precision&93.90&85.94&100.0&85.51&96.77&98.33&84.51&93.44&100.0&90.32&100.0&98.28&93.75&\multirow{2}{*}{\textbf{94.17}}\\
		&recall&94.44&91.67&73.33&98.33&100.0&98.33&100.0&95.0&98.33&93.33&90.0&95.0&100.0\\
		\midrule
		\multirow{2}{*}{Untaught} &precision&91.20&88.24&88.33&95.83&85.07&98.18&87.23&73.53&98.08&96.55&83.33&100.0&100.0&\multirow{2}{*}{\textbf{88.58}}\\
		&recall&86.11&100.0&88.33&76.67&95.0&90.0&68.33&83.33&85.0&93.33&83.33&96.67&73.33\\
		\bottomrule	
	\end{tabular}
	\label{f1score}
\end{table*}
\subsubsection{Detection Matrix}
We summarized the gesture detection results of our real-time system. Since we did not aim to evaluate the classification performance here, we depicted the detection results in Table \ref{falsetable} considering all four test subjects. Our system correctly detected 1388 true positive gestures, and provoked 25 false alarms among the total of 1864 test samples in which there are 1440 true gestures and 424 true negative RMs, respectively. Furthermore, we define two different types of miss-detections (MDs), in which the MDs from HAD means that our HAD miss-detects a gesture, while the MDs from the classifier means that, the HAD detects the gesture, but this gesture is incorrectly rejected by our classifier as a RM. The false alarm rate (FAR) and miss-detection rate (MDR) of our system are $5.90\%$ and $3.61\%$, respectively.
\begin{table}[h]
	\caption{Gesture detection matrix based on four test subjects}
	\label{falsetable}	
	\centering	
	\setlength{\tabcolsep}{0.25em}
	\begin{tabular}{c|c|c|c}
		\toprule
		Detected Positive&\multicolumn{2}{c|}{Detected Negative}& Overall\\	
		\midrule
		1388 &26 &26&1440\\
		%\hline
		True Positives&MDs from HAD&MDs from classifier&True Gestures\\
		%\cline{1-4}
		\midrule
		25&\multicolumn{2}{c|}{399}&424\\
		False Alarms&\multicolumn{2}{c|}{True Negatives}&Negative Samples\\	
		\bottomrule	
	\end{tabular}
\end{table}
\subsubsection{Runtime}
As depicted in Table \ref{runtime}, in the HIL context, we also noted the avg. runtime of the multi-feature encoder, HAD and CNN based on all the 1838 classifications, which include 1388 true positives, 399 true negatives, 25 false alarms and 26 MDs from the classifier. The multi-feature encoder includes the 2-D FFT, 25 points selection, RD and AoA estimation. It needs to be mentioned that the multi-feature encoder and the HAD were executed in the CPU using unoptimized C/C++ code, while the CNN ran in the GPU based on TensorRT. The multi-feature encoder and HAD took only approx. 7.12 ms and 0.38 ms without using any FFT acceleration engine, while the CNN took only 25.84 ms on average. The overall runtime of our proposed radar-based gesture recognition system is only approx. 33 ms.
\begin{table}[!htbp]
\caption{Average runtime of the gesture recognition system}
\label{runtime}	
\centering	
\setlength{\tabcolsep}{0.25em}
\begin{tabular}{cccc|c}
	\toprule
	%\cline{1-4}
	&Multi-feature encoder&HAD&CNN&Overall\\
	&(CPU)&(CPU)&(GPU)&\\
	\midrule
	avg. runtime&7.12 ms&0.38 ms&25.84 ms&33.15 ms\\
	\toprule		
\end{tabular}
\end{table}

\section{Conclusion}
\label{conclusion}
We developed a real-time radar-based gesture recognition system built in an edge-computing platform. The proposed multi-feature encoder could effectively encode the gesture profile, i.e., range, Doppler, azimuth, elevation, temporal information as a feature cube, which is then fed into a shallow CNN for gesture classification. Furthermore, to reduce the latency caused by the fixed number of required measurement-cycles in our system, we proposed the STA/LTA-based gesture detector, which detects the tail of a gesture. In the off-line case, based on LOOCV, our proposed gesture recognition approach achieves 98.47\% and 93.11\% avg. accuracy using gestures from taught and untaught subjects, respectively. In addition, the trained shallow CNN has a small model size and requires few GFLOPs. In the HIL context, our approach achieves 94.17\% and 88.58\% $F_1$-scores based on two taught and two untaught subjects as test sets, respectively. Finally, our system could be built in the edge-computing platform, and requires only approx. 33 ms to recognize a gesture. Thanks to the promising recognition performance and low computational complexity, our proposed radar-based gesture recognition system has the potential to be utilized for numerous applications, such as mobile and wearable devices. In future works, different gesture datasets with large diversity need to be constructed according to specific use cases. What's more, in some use cases where the radar is not stationary to the user, the classification accuracy of the proposed system might decrease and accordingly algorithms, such as ego motion compensation, could be considered.

%\clearpage

% or


%\appendices

\section*{Acknowledgment}
The authors would like to thank the editor and anonymous reviewers for giving us fruitful suggestions, which significantly improve the quality of this paper. Many thanks to the students for helping us collect the gesture dataset in this interesting work. 

%\ifCLASSOPTIONcaptionsoff

%\fi



%\bibliographystyle{IEEEtran}
\bibliographystyle{IEEEtran}
\bibliography{IEEEabrv,refs}

\begin{IEEEbiography}[{\includegraphics[width=1in,height=1.25in,clip,keepaspectratio]{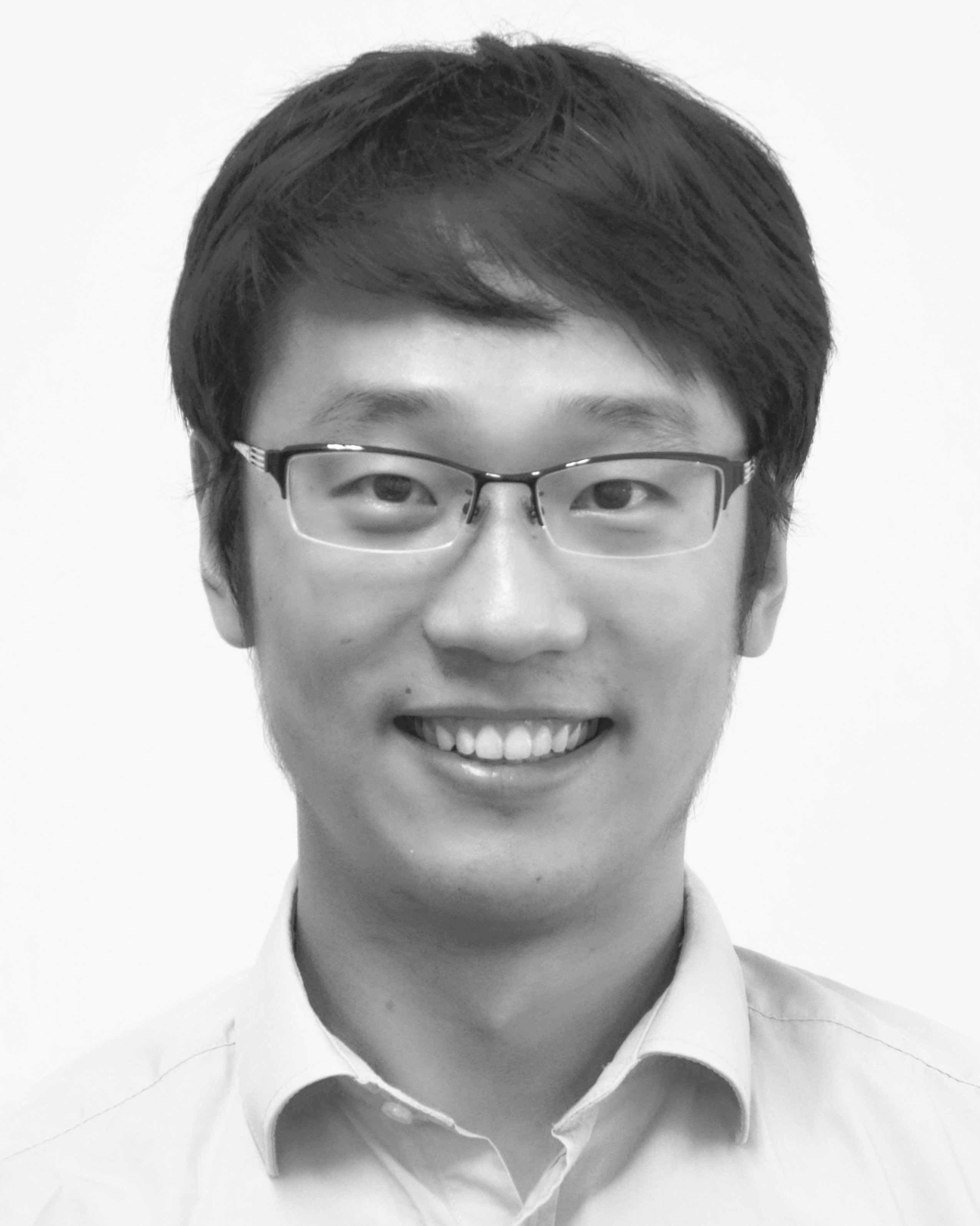}}]{Yuliang Sun} (GSM'18) received the B.Eng. degree in mechatronic engineering from Tongji University, Shanghai, China, and Aschaffenburg University of Applied Sciences, Germany, in 2014, and the M.Sc. degree in electrical and computer engineering from Technical University of Munich, Germany, in 2017. He is currently pursuing the Dr.-Ing. degree in electrical engineering with the Department of Integrated Systems, Ruhr University Bochum, Germany and the Research Institute for Automotive Electronics (E-LAB) in collaboration with HELLA~GmbH~\&~Co.~KGaA, Lippstadt, Germany. His research interests are automotive radar signal processing, radar-based human motion recognition and machine learning.

He received the Best Student Paper Award at IEEE International Radar Conference in 2020.
\end{IEEEbiography}
	
\begin{IEEEbiography}[{\includegraphics[width=1in,height=1.25in,clip,keepaspectratio]{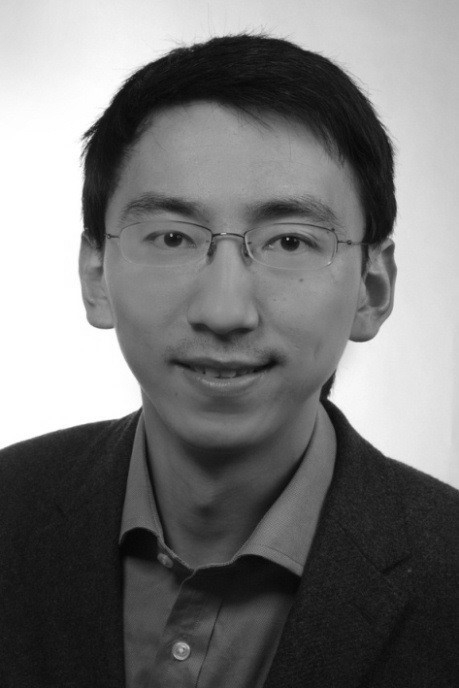}}\vfil]{Tai Fei} (S'12-M'19) received the B.Eng. degree in telecommunication engineering from Shanghai Maritime University, Shanghai, China, in 2005 and the Dipl.-Ing. and Dr.-Ing. degrees in electrical engineering and information technology from Darmstadt University of Technology (TUD), Darmstadt, Germany, in 2009 and 2014, respectively. From 2009 to 2012, he worked as a Research Associate with the Institute of Water-Acoustics, Sonar-Engineering and Signal-Theory at Hochschule Bremen, Bremen, Germany, in collaboration with the Signal Processing Group at TUD, Darmstadt, Germany, where his research interest was the detection and classification of underwater mines in sonar imagery. From 2013 to 2014, he worked as a Research Associate with Center for Marine Environmental Sciences at University of Bremen, Bremen, Germany. 
	
Since 2014, he has been working as a development engineer at HELLA GmbH \& Co. KGaA, Lippstadt, Germany, where he is mainly responsible for the development of reliable signal processing algorithms for automotive radar systems.
\end{IEEEbiography}
	
\begin{IEEEbiography}[{\includegraphics[width=1in,height=1.25in,clip,keepaspectratio]{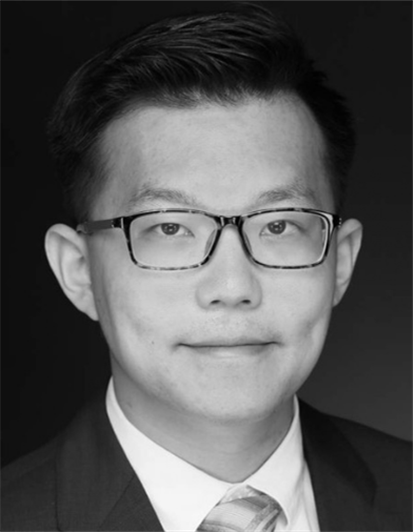}}]{Xibo Li} received the B.Sc. degree in mechanical engineering from Beijing Institute of Technology. Beijing, China. He is currently pursuing the M.Sc. degree in automotive engineering and transport at RWTH Aachen University, Aachen, Germany. 
	
	His current research interests include automotive radar signal processing, machine learning and sensor fusion. 
\end{IEEEbiography}

\begin{IEEEbiography}[{\includegraphics[width=1in,height=1.25in,clip,keepaspectratio]{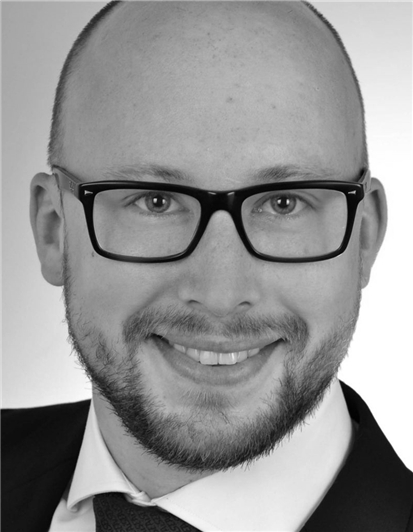}}]{Alexander Warnecke} received the M.Sc. degree in electrical engineering and the Dr.-Ing. degree in degradation mechanisms NMC based lithium-ion batteries from RWTH Aachen University, Aachen, Germany, in 2012 and 2017, respectively. As a Research Associate at the Institute for Power Electronic and Electrical Drives (ISEA), he was involved in several projects related to ageing of lithiumion batteries at the chair for electrochemical energy conversion and storage systems. From 2015 to 2017, he was an Executive Manager at the Center for Ageing, Reliability, and Lifetime Prediction (CARL). 
	
	He is currently the head of the Research Institute for Automotive Electronics (E-LAB), HELLA~GmbH~\&~Co.~KGaA, Lippstadt, Germany.
\end{IEEEbiography}

\begin{IEEEbiography}[{\includegraphics[width=1in,height=1.25in,clip,keepaspectratio]{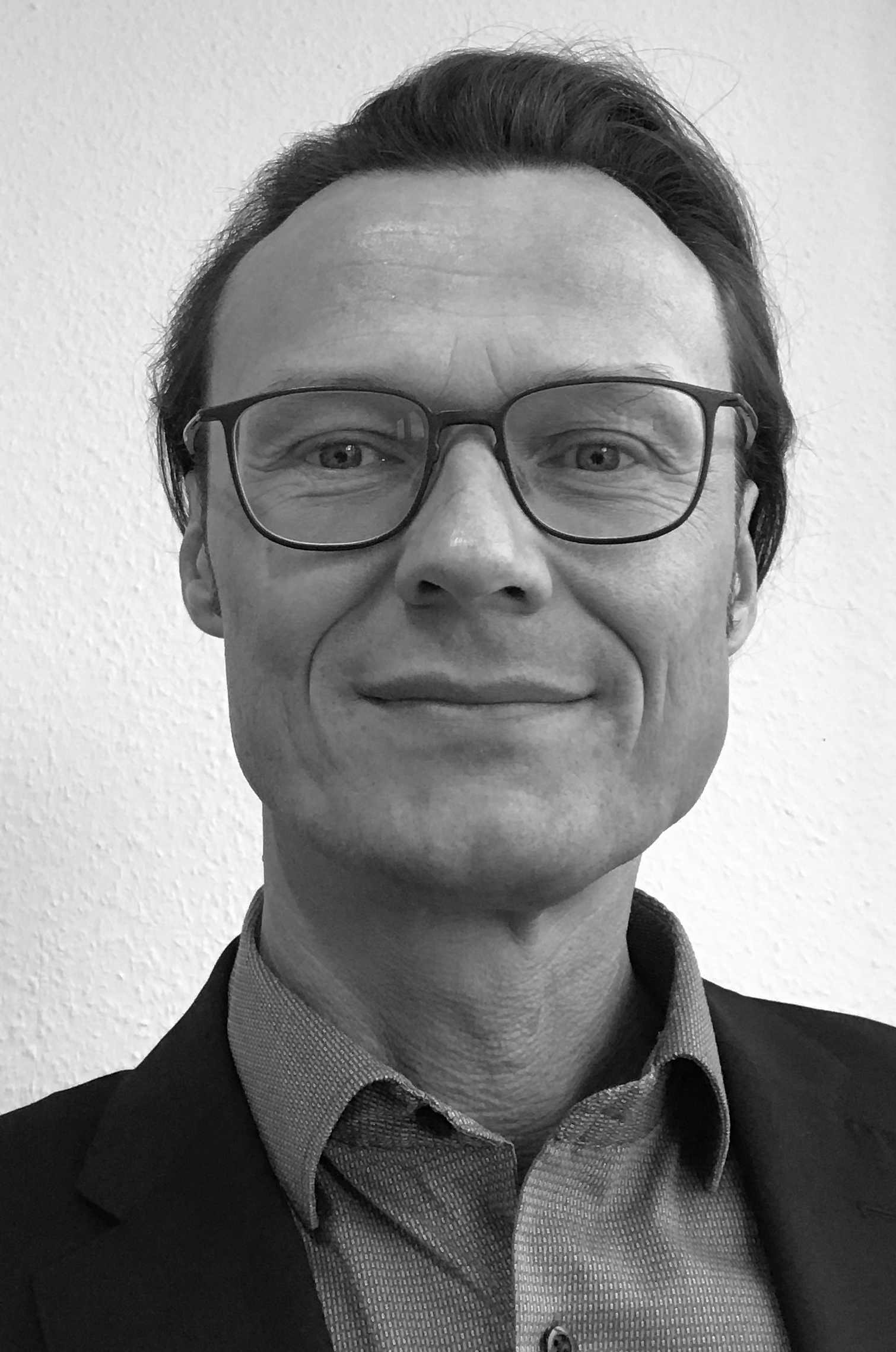}}]{Ernst Warsitz} received the Dipl.-Ing. and Dr.-Ing. degrees in electrical engineering from Paderborn University, Paderborn, Germany, in 2000 and 2008, respectively.	He joined the Department of Communications Engineering of the University of Paderborn in 2001 as a Research Staff Member, where he was involved in several projects related to single- and multi-channel speech processing and automated speech recognition. From 2007 he worked as a development engineer at HELLA GmbH \& Co. KGaA, Lippstadt, Germany, in the field of signal processing algorithms for automotive radar systems. He is currently the head of the Radar Signal Processing and Signal Validation Department at HELLA GmbH \& Co. KGaA, Lippstadt, Germany.
	
\end{IEEEbiography}

\begin{IEEEbiography}[{\includegraphics[width=1in,height=1.25in,clip,keepaspectratio]{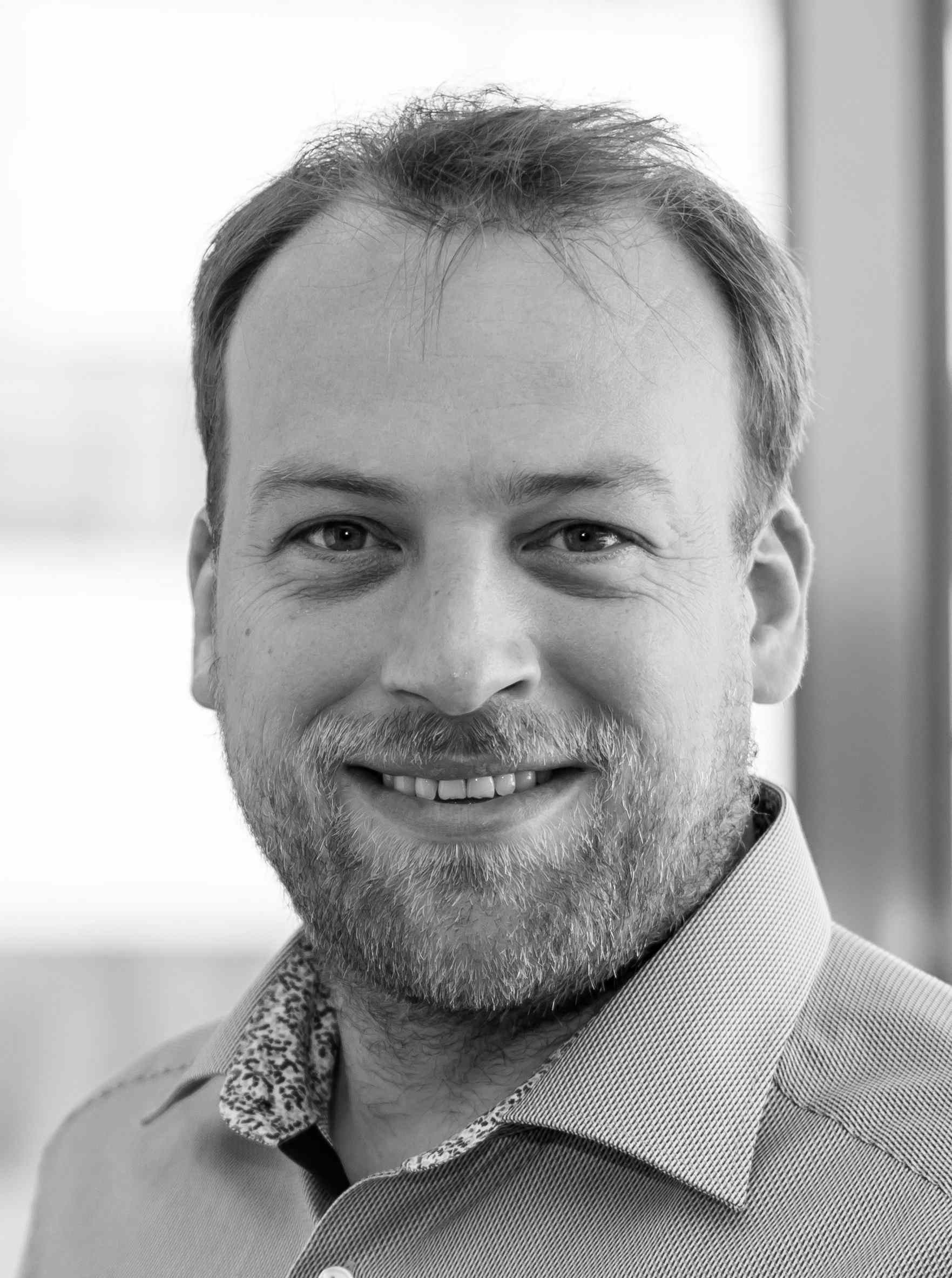}}]{Nils Pohl} (GSM'07-M'11-SM'14) received the Dipl.-Ing. and Dr.-Ing. degrees in electrical engineering from Ruhr University Bochum, Bochum, Germany, in 2005 and 2010, respectively. 
	From 2006 to 2011, he was a Research Assistant with Ruhr University Bochum, where he was involved in integrated circuits for millimeter-wave (mm-wave) radar applications. In 2011, he became an Assistant Professor with Ruhr University Bochum. In 2013, he became the Head of the Department of mm-wave Radar and High Frequency Sensors with the Fraunhofer Institute for High Frequency Physics and Radar Techniques, Wachtberg, Germany. In 2016, he became a Full Professor for Integrated Systems with Ruhr University Bochum. He has authored or coauthored more than 100 scientific papers and has issued several patents. His current research interests include ultra-wideband mm-wave radar, design, and optimization of mm-wave integrated SiGe circuits and system concepts with frequencies up to 300 GHz and above, as well as frequency synthesis and antennas.
	
	Prof. Pohl is a member of VDE, ITG, EUMA, and URSI. He was a co-recipient of the 2009 EEEfCom Innovation Award, the 2012 EuMIC Prize, and the 2015 Best Demo Award of the IEEE Radio Wireless Week, and a recipient of the Karl-Arnold Award of the North Rhine-Westphalian Academy of Sciences, Humanities and the Arts in 2013 and the IEEE MTT Outstanding Young Engineer Award in 2018.
\end{IEEEbiography}

\end{document}